# $\beta$-PtO$_2$: Phononic, thermodynamic, and elastic properties derived from first-principles calculations

Quan Chen (陈泉)[1,2], Wei Li (李伟)[1], Yong Yang (杨勇)[1,3,†]

1 *Key Laboratory of Materials Physics, Institute of Solid State Physics, Chinese Academy of Sciences, Hefei 230031, China*

2 *Institutes of Physical Science and Information Technology, Anhui University, Hefei 230601, China*

3 *Science Island Branch of Graduate School, University of Science and Technology of China, Hefei 230026, China*

Corresponding author. E-mail: [†]yyanglab@issp.ac.cn

*β*-PtO$_2$ is a useful transition metal dioxide, but its fundamental thermodynamic and elastic properties remain unexplored. Using first-principles calculations, we systematically studied the structure, phonon, thermodynamic and elastic properties of *β*-PtO$_2$. The lattice dynamics and structural stability of *β*-PtO$_2$ under pressure were studied using the phonon spectra and vibrational density of states. The vibrational frequencies of the optical modes of *β*-PtO$_2$ increase with elevating pressure; this result is comparable with the available experimental data. Then, the heat capacities and their pressure responses were determined based on the phonon calculations. The pressure dependence of the Debye temperature was studied, and the results are compared in two distinct aspects. The elastic moduli of *β*-PtO$_2$ were estimated through the Voigt–Reuss–Hill approximation. The bulk modulus of *β*-PtO$_2$ increases linearly with pressure, but the shear modulus is nearly independent of pressure. Our study revealed that the elastic stiffness coefficients $C_{44}$, $C_{55}$ and $C_{66}$ play a primary role in the slow variation of the shear modulus.

## 1 Introduction

In recent years, transition metal oxides have been extensively studied owing to their unique electronic structures and thermodynamic and elastic properties [1–3]. Among these compounds, $PtO_2$ has received particular attention as a general-purpose catalyst with wide applications in optics, electrochemistry, and organic syntheses [4–6]. The rich body of experimental works on $PtO_2$, such as the synthesis of water-oxidizing catalysts [7] and the reaction behavior of $PtO_2$ in hydrosilylation [8], is mainly focused on practical applications. To date, the lattice dynamics and thermodynamic properties of $PtO_2$ are little known, although the electronic properties have been well reported [9–11]. Experimental studies have determined three crystalline phases of $PtO_2$ [12–14]: $α$-$PtO_2$, $β$-$PtO_2$, and $β'$-$PtO_2$, which have hexagonal $CdI_2$-, orthorhombic $CaCl_2$-, and tetragonal rutile-type structures, respectively. Unlike other Group VIII metal dioxides, all of which have a rutile structure [15], $PtO_2$ is the most stable in the beta phase ($β$-$PtO_2$) under ambient conditions. Based on the results of density of states, Wu and Weber concluded that this anomaly originates from strong hybridization of the Pt 5d and O 2p states at the Fermi level [9]. A comprehensive knowledge of the fundamental properties of $PtO_2$ is a prerequisite for elucidating the role and microscopic mechanism of $PtO_2$ in applications in fields such as optics and catalysis.

Motivated by the lack of basic properties in the existing literature, we systematically studied the phononic, thermodynamic, and elastic properties of $β$-$PtO_2$ through first-principles calculations. The specific heat, Debye temperature, and elastic modulus were determined for the first time, and the optical vibrational modes of $β$-$PtO_2$ are comprehensively investigated. Our calculations show that $β$-$PtO_2$ is stable over a wide pressure range. Increasing the applied pressure increases the vibrational frequencies of $β$-$PtO_2$ and decreases the heat capacity by hardening the atomic vibrations. The macroscopic effective elastic constants are deduced from the phononic and specific-heat calculations and compared with those derived from the single-crystal elastic stiffness. The calculated effective shear moduli of polycrystalline systems of $β$-$PtO_2$ are nearly pressure-independent owing to the slow variation of the elastic stiffness coefficients $C_{44}$, $C_{55}$, and $C_{66}$.

## 2 Computational methods

The structural optimization and total energy calculations were performed using the Vienna ab-initio Simulation Package (VASP) [16, 17], which is based on density functional theory (DFT).

The exchange-correlation interactions of electrons were described using the local density approximation (LDA) [18, 19] and compared with the results of the generalized gradient approximation (GGA) using the Perdew–Burke–Erzerholf (PBE)-type functional [20]. Using a plane wave basis set the Kohn-Sham equations were solved, with the projected augmented wave (PAW) method [21] to describe the electron-ion core interactions. Energy cutoff of the plane-wave basis set is 600 eV. The Brillouin zone (BZ) was sampled on a 16 × 16 × 16 Monkhorst–Pack k-mesh [22]. The vibrational and thermodynamic properties of $\beta$-PtO$_2$ were studied by combining the density functional perturbation theory (DFPT) as implemented in VASP with the direct method implemented in PHONOPY package [23]. Phonon calculations were performed in a 2 × 2 × 2 supercell on a 6 × 6 × 6 k-mesh. The lattice constants and phonons were calculated by both LDA and GGA for comparison. Additionally, the reliability of the standard DFT–GGA calculations was ensured via two methods: the first is the GGA + U method [24], and the second method involves hard pseudopotentials. The hard pseudopotentials were constructed in the pseudo-atomic configurations $5p^65d^96s^1$ and $2s^22p^4$ for Pt and O, respectively, with the plane-wave energy cutoff set to 950 eV. In the GGA + U calculation, the effective on-site Coulomb repulsion ($U_{eff}$) of the 5d orbitals of Pt was set to 7.5 eV [25]. The thermodynamic and elastic properties were calculated via only GGA.

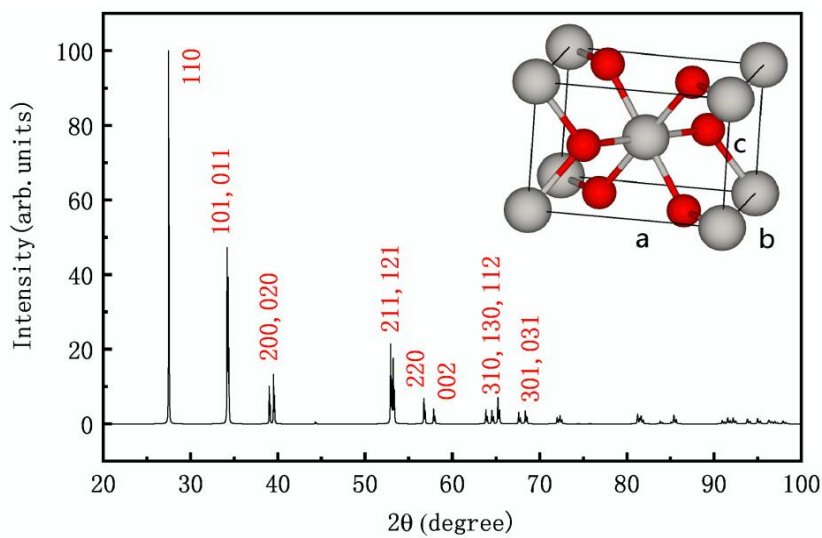

**Fig. 1** Crystal structure and simulated X-ray diffraction pattern of β-PtO$_2$. The large gray and small red balls represent Pt and O atoms, respectively.

## 3. Results and discussion

## 3.1. Structural properties

Figure 1 shows the crystal structure and simulated X-ray diffraction pattern of $\beta$-$PtO_2$. The results indicate that $\beta$-$PtO_2$ has an orthorhombic $CaCl_2$-type crystal structure, which is similar to the rutile structure but with a slight distortion in the directions of the cell axes *a* and *b*. Two Pt atoms and six O atoms are located in the *2a* and *4g* Wyckoff positions, respectively. The Pt atoms are surrounded by four and two oxygen atoms at 2.043 Å and 2.036 Å, respectively. The strongest signal in the X-ray diffraction data was found at $2\theta = 27.49°$, with a *d* spacing of 3.24 Å. The corresponding diffraction plane is (110). The second strongest diffraction appears in the (101) and (011) planes in the 34.19°–34.41° and 2.61–2.62 Å ranges of incident angles and *d*-spacing, respectively.

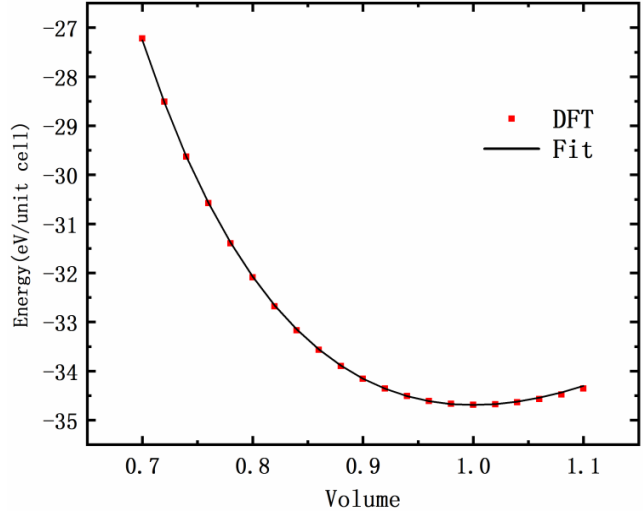

**Fig. 2** Total energy of $\beta$-$PtO_2$ as a function of volume (normalized to the equilibrium volume at $P = 0$). The solid curve is the data fitted according to the Murnaghan equation of state.

We calculated the total energy *E* by changing the lattice volume *V* to get a series of *E-V* data, and then the data are least squares fitted the Murnaghan equation of state [26] to obtain the elastic properties:

$$E_{tot}(V) = \frac{B_0 V}{B_0'} \left[ \frac{(V_0/V)^{B_0'}}{B_0' - 1} + 1 \right] + E_0 - \frac{B_0 V_0}{B_0' - 1} \quad (1),$$

where $E_0$ and $V_0$ are the total energy and volume at equilibrium, while $B_0$ and $B_0'$ are the bulk modulus and its pressure derivative, respectively. The results are shown in Fig. 2, and the calculated data are listed in Table 1. The calculated lattice constants are in good agreement with the experimental and other theoretical data [15, 27]. The results calculated through the GGA + U computation differ from that calculated through the standard method by less than 5%, indicating that the choice of the Hubbard term $U$ does not significantly affect the main properties of $\beta$-PtO$_2$. The deduced values of $B_0$ and $B_0'$ are 209 GPa and 4.06, respectively. To our knowledge, the first set of data on the bulk modulus of β-PtO$_2$ is reported in this paper. The pressure was determined using the Murnaghan equation of state:

$$P = -\left(\frac{\partial E}{\partial V}\right) = (B_0/B_0')[(V_0/V)^{B_0'} - 1] \qquad (2),$$

Accordingly, we can obtain the volume and energy at any pressure point within a certain range, e.g., 0-30 GPa in our study

**Table 1** Calculated lattice constants, volumes, bulk moduli, and pressure derivatives of the bulk modulus.

|  | $a$ (Å) | $b$ (Å) | $c$ (Å) | $V_0$ (Å$^3$) | $B_0$ (GPa) | $B_0'$ |
|---|---|---|---|---|---|---|
| GGA | 4.613 | 4.557 | 3.185 | 66.95 | 209 | 4.06 |
| LDA | 4.492 | 4.442 | 3.142 | 62.69 | 265 | 3.90 |
| GGA + U | 4.557 | 4.569 | 3.136 | 65.30 | 220 | 3.81 |
| Experiment [15] | 4.484 | 4.539 | 3.136 | 63.83 |  |  |
| Other works [27] | 4.61 | 4.55 | 3.19 | 66.91 |  |  |

**3.2 Phonon properties**

Figure 3 illustrates the optical vibrational modes of $\beta$-PtO$_2$ calculated via the DFPT method at the gamma point. For simplicity, we refer to these modes as ω1–ω15, where the frequency ranges from high (ω1) to low (ω15). Note that ω2, ω4, ω7, ω8, ω10, and ω12 represent the vibrations of O atoms only, which are Raman-active modes, and consistent with the experimentally measured Raman spectra [28]. As the other modes of $\beta$-PtO$_2$ lack Raman activity, they must be measured via other detection methods, such as infrared spectroscopy. Table 2 compares the calculated frequencies of the optical modes of $\beta$-PtO$_2$ and the experimental data. In general, the frequencies were higher in the LDA calculation than in the GGA calculation; therefore, the LDA values better matched the experimental data in the high-frequency region. However, the frequencies calculated through GGA were closer to the experimental values in the low-frequency region. The phonon modes of $\beta$-PtO$_2$ can be roughly divided into three categories. The high-frequency modes of $\beta$-

PtO$_2$, ω1–ω8 (500 < ω < 800 cm$^{-1}$), are attributed to the coupled modes that mainly involve the stretching of the Pt–O bonds. The mid-frequency modes of β-PtO$_2$, ω9–ω12 (250 < ω < 450 cm$^{-1}$), correspond to the O–Pt–O bending vibrations in the PtO$_6$ octahedron. Finally, the low-frequency modes of β-PtO$_2$, ω13–ω15 (ω < 250 cm$^{-1}$), originate from the interactions between different Pt polyhedrons.

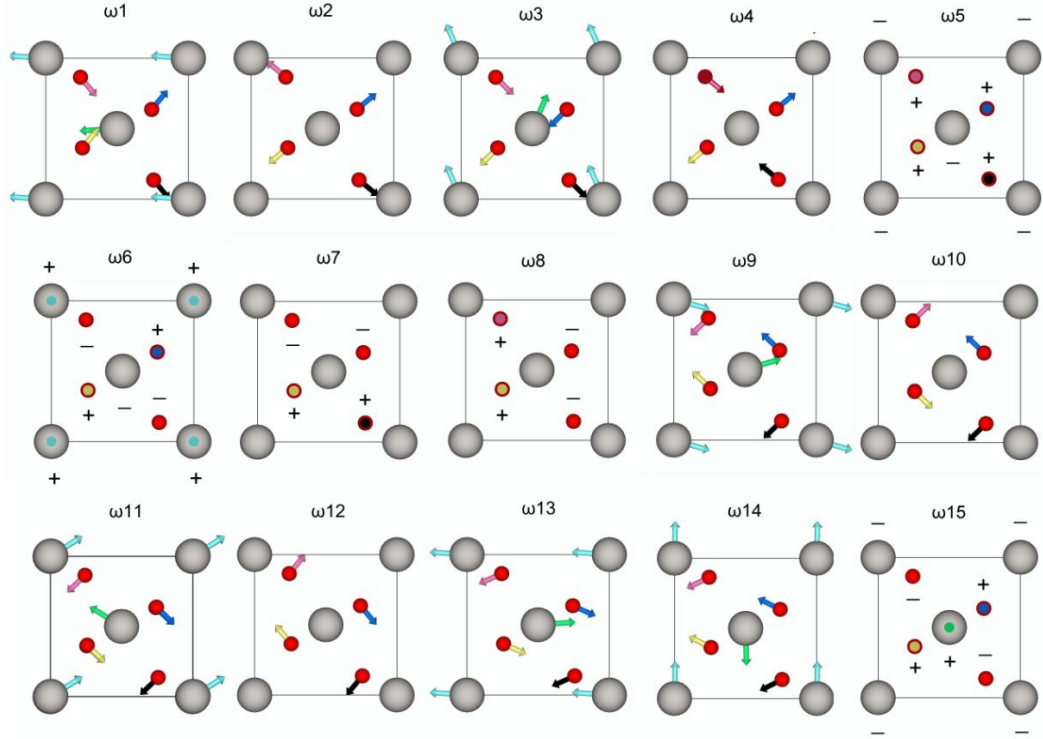

**Fig. 3** Top views of eigenvectors of the optical vibrational modes of β-PtO$_2$. The gray and red balls represent the Pt and O atoms, respectively. The symbols + and − indicate that the vibration directions are upward and downward, respectively, along the c axis.

To study the atomic-scale dynamical properties, we calculated the phonon dispersion curve and vibrational density of states (VDOS) of β-PtO$_2$ along some high-symmetry points through both LDA and GGA for pressures up to 30 GPa. The results are displayed in Figure 4. The phonon spectra and VDOS indicate that β-PtO$_2$ is dynamically stable over the pressure range 0–30 GPa. Increasing the pressure shifted all optical phonon modes of β-PtO$_2$ upward (toward higher frequencies). The GGA better described the previous experimentally measured vibrational frequencies in the acoustic and low-frequency optical phonon branches, and the LDA better described the high-frequency optical branches. In addition, the gaps between some dispersion bands in the high-frequency region were enlarged under pressure, indicating the different pressure responses of the frequency modes of β-PtO$_2$.

**Table 2** Calculated frequencies of the optical vibrational modes of $\beta$-PtO$_2$ and (for comparison) the available experimental data.

| Mode | Frequency (cm$^{-1}$) | | | Symmetry (D$_{2h}$) |
|---|---|---|---|---|
| | LDA | GGA | Expt[a] | |
| $\omega$1 | 728 | 665 | | |
| $\omega$2 | 716 | 651 | (~790) | **B$_{1g}$** |
| $\omega$3 | 700 | 635 | | |
| $\omega$4 | 697 | 627 | 743 | **A$_g$** |
| $\omega$5 | 683 | 626 | | |
| $\omega$6 | 677 | 624 | | |
| $\omega$7 | 582 | 532 | 616 | **B$_{3g}$** |
| $\omega$8 | 561 | 512 | 606 | **B$_{2g}$** |
| $\omega$9 | 402 | 383 | | |
| $\omega$10 | 391 | 371 | 340 | **B$_{1g}$** |
| $\omega$11 | 378 | 358 | | |
| $\omega$12 | 279 | 254 | 205 | **A$_g$** |
| $\omega$13 | 223 | 208 | | |
| $\omega$14 | 222 | 202 | | |
| $\omega$15 | 183 | 167 | | |

[a]Reference [28]

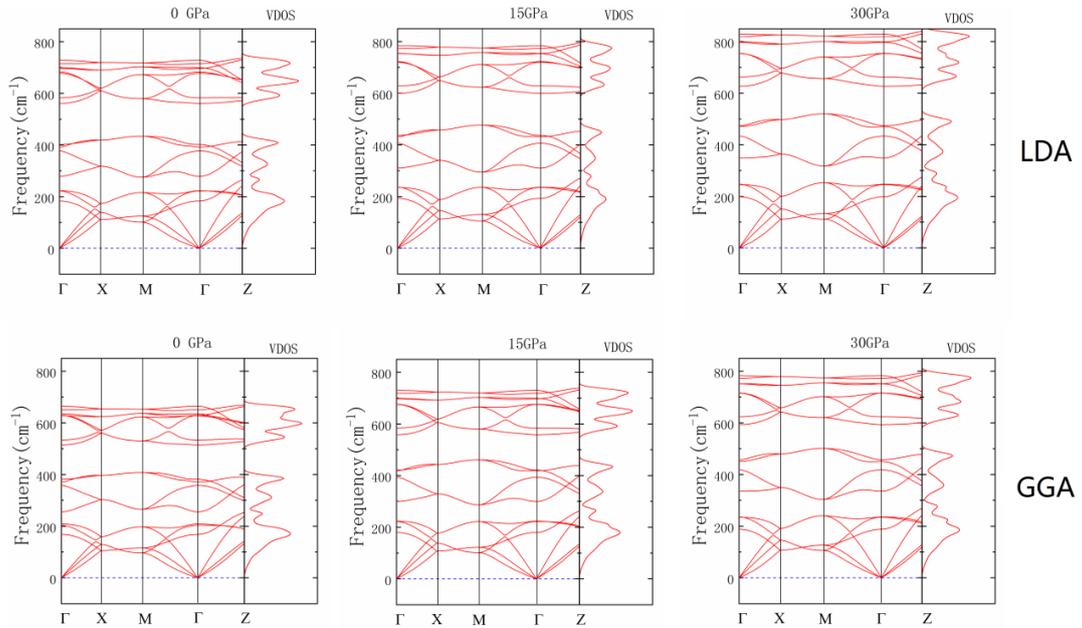

**Fig. 4** Calculated phonon dispersions along some high-symmetry lines of the BZ, and the VDOS of $\beta$-PtO$_2$ under different pressures.

## 3.3 Thermodynamic properties

The heat capacity (specific heat in the case of unit volume, mass, or number of particles) is a basic thermodynamic quantity that describes the extent to which a substance absorbs and releases heat. From a microscopic viewpoint, the heat capacity of a substance manifests owing to the thermal motions of atoms and electrons. In semiconductors and insulators, the main contribution to heat capacity is atomic vibrations, i.e., phonons. Based on first-principles phonon calculations and statistical mechanics, the constant-volume heat capacity is given as follows:

$$C_V = \left(\frac{\partial E}{\partial T}\right)_V = \sum_{qv} k_B \left(\frac{\hbar\omega(qv)}{k_B T}\right)^2 \frac{exp(\hbar\omega(qv)/k_B T)}{[exp(\hbar\omega(qv)/k_B T)-1]^2}, \quad (3)$$

where $\hbar$ is the reduced Planck constant, $k_B$ is the Boltzmann constant, $T$, $\omega$, and $q$ denote the temperature, frequency and wave vector, respectively, and the index $v$ distinguishes the various optical phonon branches. $C_v$ is well approximated using the Debye model, which describes the change in heat capacity with temperature. The general form of the Debye model is as follows [29]

$$C_V = 9N_A k_B \left(\frac{T}{\Theta_D}\right)^3 \int_0^{\Theta_D/T} \frac{x^4 e^x}{(e^x-1)^2} dx, \quad (4)$$

where $N_A$ is the Avogadro constant, $\Theta_D$ is the Debye temperature, and the other parameters have their usual meanings. In the low-temperature limit ($T \rightarrow 0$ when $\Theta_D/T \rightarrow \infty$), the Debye model reduces $C_V$ to $\sim \frac{12Nk_B\pi^4}{5}\left(\frac{T}{\Theta_D}\right)^3$ [29, 30], which is proportional to $T^3$; this is called the Debye $T^3$ law. To verify the applicable range of this law, we compared the specific heat obtained through DFT calculations (which are particularly accurate at room temperature and below) with the curve fitted to the Debye model (see Fig. 5). By fitting the specific-heat curve to Eq. (3), we estimated the Debye temperature of $\beta$-PtO$_2$ as ~512 K.

At temperatures below 20 K, the two curves are almost identical. Above 20 K, the Debye model deviates significantly from the specific heat predicted by first-principles calculations. Therefore, the Debye T$^3$ law applies when $T \leq \frac{1}{30}\Theta_D$. In fact, the value of the integral $\int_0^{30} \frac{x^4 e^x}{(e^x-1)^2} dx$ in Eq. (4) differs by only ~1.5×10$^{-8}$ from the exact value ($\int_0^\infty \frac{x^4 e^x}{(e^x-1)^2} dx = \frac{4\pi^4}{15}$ at the low temperature limit), justifying the aforementioned temperatures at which the Debye model can be applied ($T \leq \frac{1}{30}\Theta_D$). On the other hand, in the high temperature region, the calculated specific heat approaches the classical thermodynamic limit of 3$R$ [31].

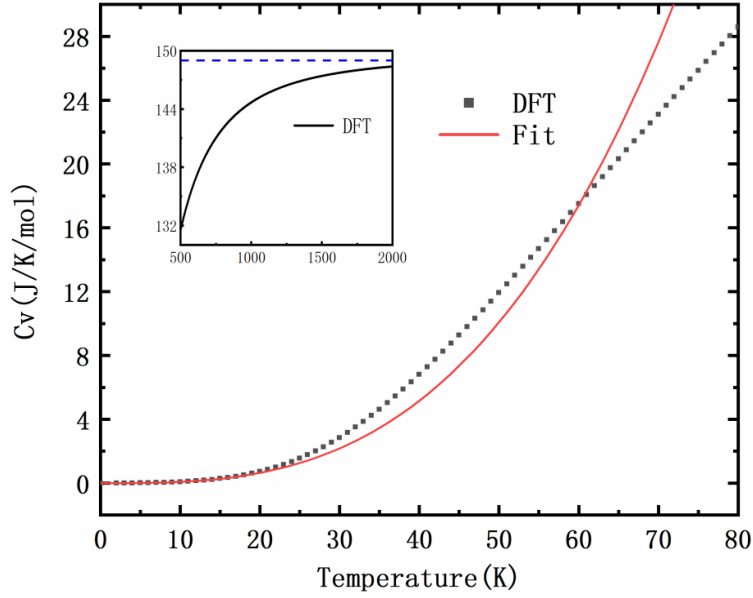

**Fig. 5** Calculated specific heat of $\beta$-PtO$_2$ at low temperatures, and (for comparison) the heat capacity fitted using the Debye model. The inset shows the high-temperature behavior of the specific heat.

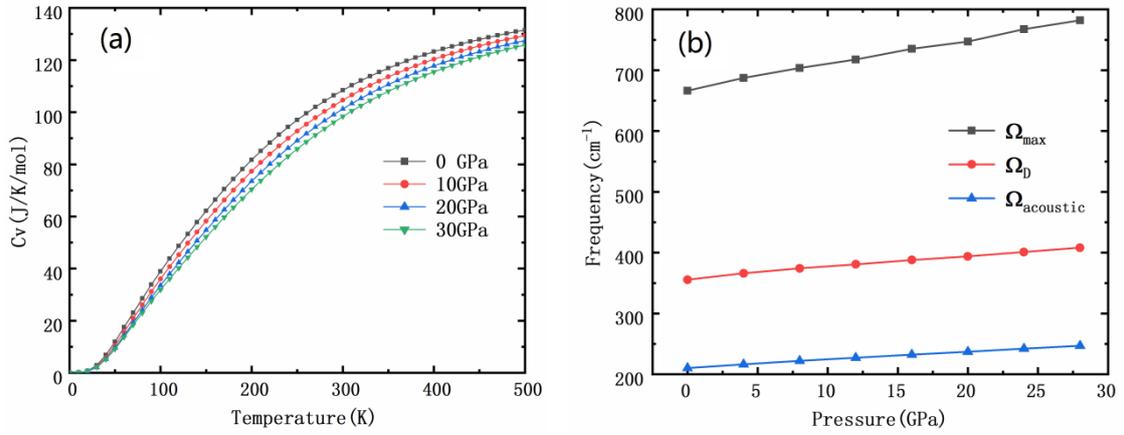

**Fig. 6 (a)** Calculated heat capacity of $\beta$-PtO$_2$ under different pressures. **(b)** Maximum phonon frequency of the gamma point, Debye frequency, and maximum acoustic frequency as functions of pressure.

Figure 6(a) plots the temperature dependence of the specific heat and its variations with pressure. The specific heat gradually increases and decreases with increasing temperature and pressure, respectively; this occurs because the compression of crystal volume enhances the lattice vibrations and increases the vibrational frequencies, thus reducing the heat absorption in the same vibrational mode. Figure 6(b) shows the pressure dependence of some typical vibrational frequencies. The maximum frequency of the optical modes at the gamma point, Debye frequency, and maximum frequency of the acoustic branches monotonically increase with pressure because the interatomic force constants increase with pressure. The Debye frequency is located between the maximum frequency of the gamma point (i.e., the maximum frequency of the optical mode of $\beta$-PtO$_2$ in the long wavelength limit) and that of the acoustic branch (deduced from the VDOS). The numerical values of these three frequencies are quite different, indicating that these two frequencies cannot simply replace the Debye frequency in the Debye model. Moreover, the Debye frequency is almost proportional to the acoustic branch frequency, which is consistent with the fact that the acoustic phonons are the main contributors to the low-temperature specific heat.

**Table 3** Calculated elastic stiffness coefficients $C_{ij}$ (GPa), bulk modulus $B$ (GPa), shear modulus $G$ (GPa), Young's modulus $E$ (GPa), and Possion's ratio $v$ of $\beta$-PtO$_2$ at different pressures.

| pressure (GPa) | $C_{11}$ | $C_{22}$ | $C_{33}$ | $C_{44}$ | $C_{55}$ | $C_{66}$ | $C_{12}$ | $C_{13}$ | $C_{23}$ | $B$ | $G$ | $E$ | $v$ |
|---|---|---|---|---|---|---|---|---|---|---|---|---|---|
| 0 | 272 | 242 | 413 | 171 | 101 | 64 | 185 | 143 | 140 | 204.5 | 85.6 | 225.0 | 0.316 |
| 4 | 286 | 254 | 431 | 180 | 102 | 63 | 201 | 155 | 150 | 217.7 | 86.3 | 228.6 | 0.325 |
| 8 | 304 | 266 | 456 | 184 | 104 | 60 | 218 | 172 | 159 | 232.8 | 86.4 | 230.7 | 0.335 |
| 12 | 328 | 280 | 469 | 190 | 105 | 57 | 230 | 187 | 170 | 246.5 | 88.4 | 236.9 | 0.340 |
| 16 | 344 | 293 | 486 | 196 | 106 | 53 | 246 | 202 | 181 | 260.7 | 88.0 | 237.3 | 0.348 |
| 20 | 365 | 306 | 502 | 199 | 107 | 50 | 260 | 218 | 190 | 274.2 | 88.6 | 240.0 | 0.354 |
| 24 | 375 | 319 | 525 | 204 | 106 | 45 | 278 | 232 | 202 | 288.9 | 86.3 | 235.5 | 0.364 |
| 30 | 415 | 344 | 541 | 208 | 109 | 44 | 292 | 254 | 216 | 308.6 | 91.0 | 248.6 | 0.366 |

**3.4 Elastic properties**

The mechanical and thermodynamic behavior of materials can be characterized by analyzing their elastic properties on the atomic scale. The elastic constants, which reflect the response of the crystal to external strain, are important for understanding the structural stability and strength of a substance. $\beta$-PtO$_2$ is structurally orthorhombic with nine independent elastic stiffness coefficients: $C_{11}$, $C_{22}$, $C_{33}$, $C_{44}$, $C_{55}$, $C_{66}$, $C_{12}$, $C_{13}$, and $C_{23}$ [32]. The effective bulk modulus $B$ and shear modulus $G$ of $\beta$-PtO$_2$ polycrystals can be obtained by combining these coefficients with the Voigt–Reuss–Hill approximation [33–35]:

$$B_V = (1/9)[C_{11} + C_{22} + C_{33} + 2(C_{12} + C_{13} + C_{23})],$$

$$G_V = (1/15)[C_{11} + C_{12} + C_{13} + 3(C_{44} + C_{55} + C_{66}) - (C_{12} + C_{13} + C_{23})],$$

$$B_R = \Delta[C_{11}(C_{22} + C_{33} - 2C_{23}) + C_{22}(C_{33} - 2C_{13}) - 2C_{33}C_{12} + C_{12}(2C_{23} - C_{12}) + C_{13}(2C_{12} - C_{13}) + C_{23}(2C_{13} - C_{23})]^{-1}$$

,

$$G_R = 15\{4[C_{11}(C_{22} + C_{33} + C_{23}) + C_{22}(C_{33} + C_{13}) + C_{33}C_{12} - C_{12}(C_{23} + C_{12}) - C_{13}(C_{12} + C_{13}) - C_{23}(C_{13} + C_{23})]/\Delta + 3[(1/C_{44}) + (1/C_{55}) + (1/C_{66})]\}^{-1},$$

$$\Delta = C_{13}(C_{12}C_{23} - C_{13}C_{22}) + C_{23}(C_{12}C_{13} - C_{23}C_{11}) + C_{33}(C_{11}C_{22} - C_{12}^2),$$

$$B = 1/2(B_R + B_V), G = 1/2(G_R + G_V), \qquad (5)$$

where the subscripts $V$ and $R$ denote the elastic moduli obtained via the methods of Voigt and Reuss, respectively. From $B$ and $G$, the Young's modulus $E$ and Poisson's ratio $v$ are obtained as follows [36]:

$$E = \frac{9BG}{3B+G}, v = \frac{3B-2G}{6B+2G}. \qquad (6)$$

To study these characteristics systematically, we calculated the elastic constants and related moduli parameters of $\beta$-PtO$_2$ at 0 K, varying the pressure from 0 to 30 GPa. The results of these calculations are listed in Table 3. The elastic stiffness coefficients increase with increasing pressure; the sole exception is $C_{66}$, which decreases with increasing pressure. Further, the bulk modulus $B$, Young's modulus $E$, and Poisson's ratio $v$ increase with increasing pressure, while the shear modulus $G$ only fluctuates slightly with pressure, which exhibits no obvious trend.

The Debye temperature, an important thermodynamic parameter, is closely related to many basic properties of solids, such as the specific heat, thermal conductivity, and melting temperature. As the vibration excitation at low temperatures is mainly due to the acoustic modes of $\beta$-PtO$_2$, the Debye temperature of solids can be calculated by averaging the sound velocity $v_m$ [37] as follows:

$$\Theta_e = \frac{h}{k_B}\left[\frac{3n}{4\pi}\left(\frac{N_A \rho}{M}\right)\right]^{1/3} v_m, \qquad (7)$$

where $h$ is the Planck constant, $n$ is the number of atoms per unit cell, , $\rho$ is the density of the material, $M$ is the relative atomic mass, and the other parameters have the usual meanings as above. In the Debye model, the average sound velocity of a material is given as follows:

$$v_m = \left[\frac{1}{3}\left(\frac{2}{v_t^3} + \frac{1}{v_l^3}\right)\right]^{-1/3}, \qquad (8)$$

where $v_l$ and $v_t$ are the longitudinal and transverse sound velocities, respectively; these can be derived from Navier's equation [29]:

$$v_l = \sqrt{\frac{3B+4G}{3\rho}}, \quad v_t = \sqrt{\frac{G}{\rho}}. \qquad (9)$$

Further, we calculated the phonon-determined Debye temperature ($\Theta_\omega$) by fitting the low-temperature specific heat to the Debye $T^3$ law and compared it with the acoustic Debye temperature ($\Theta_e$) obtained from Eq. (7). The results of this comparison are plotted in Fig. 7 as functions of pressure. Unexpectedly, the two plots were quite different. The phonon-determined Debye temperature increased almost linearly with increasing pressure, whereas the acoustic Debye temperature was relatively uniform and insensitive to pressure.

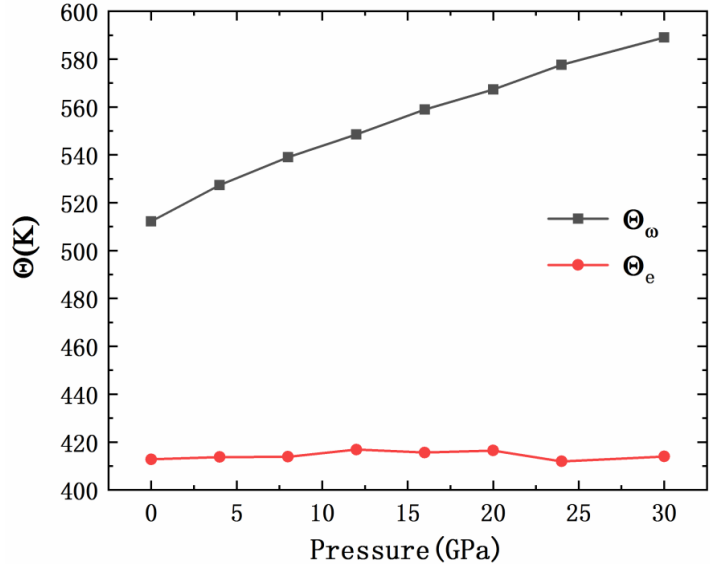

**Fig. 7** Pressure dependences of the phonon-determined and acoustic Debye temperatures ($\Theta_\omega$ and $\Theta_e$, respectively).

To find the cause of this difference, we plot the pressure responses of the bulk modulus $B$, shear modulus $G$, and sound velocity in Figure 8. The macroscopic effective bulk modulus, $B_{fit}$, was obtained by fitting the $E$-$V$ data to the Murnaghan equation of state. The macroscopic effective shear modulus, $G_{fit}$, and fitted sound velocities were derived from $B_{fit}$ and the phonon-determined Debye temperatures through Eqs. (7)–(9). Both the Voigt and Reuss approximations yield good results for the bulk modulus, but the shear moduli computed using both approximations largely deviated from $G_{fit}$. This discrepancy originates the fact that shear modulus is a macroscopic quantity, which is only applicable to isotropic systems in principle; however, the system under consideration is an orthorhombic one. Consequently, the Debye temperatures calculated by fitting the low-temperature specific heat of first-principles are suitable for single-crystal systems, and the values estimated using Eq. (7) are applicable to polycrystalline systems. Similarly, the effective elastic moduli ($B_{fit}$, $G_{fit}$) deduced from the low-temperature specific heat are only applicable to single-crystals of $\beta$-PtO$_2$, while the ones calculated using the Voigt- Reuss-Hill approximation are suitable for the polycrystalline phases of $\beta$-PtO$_2$.

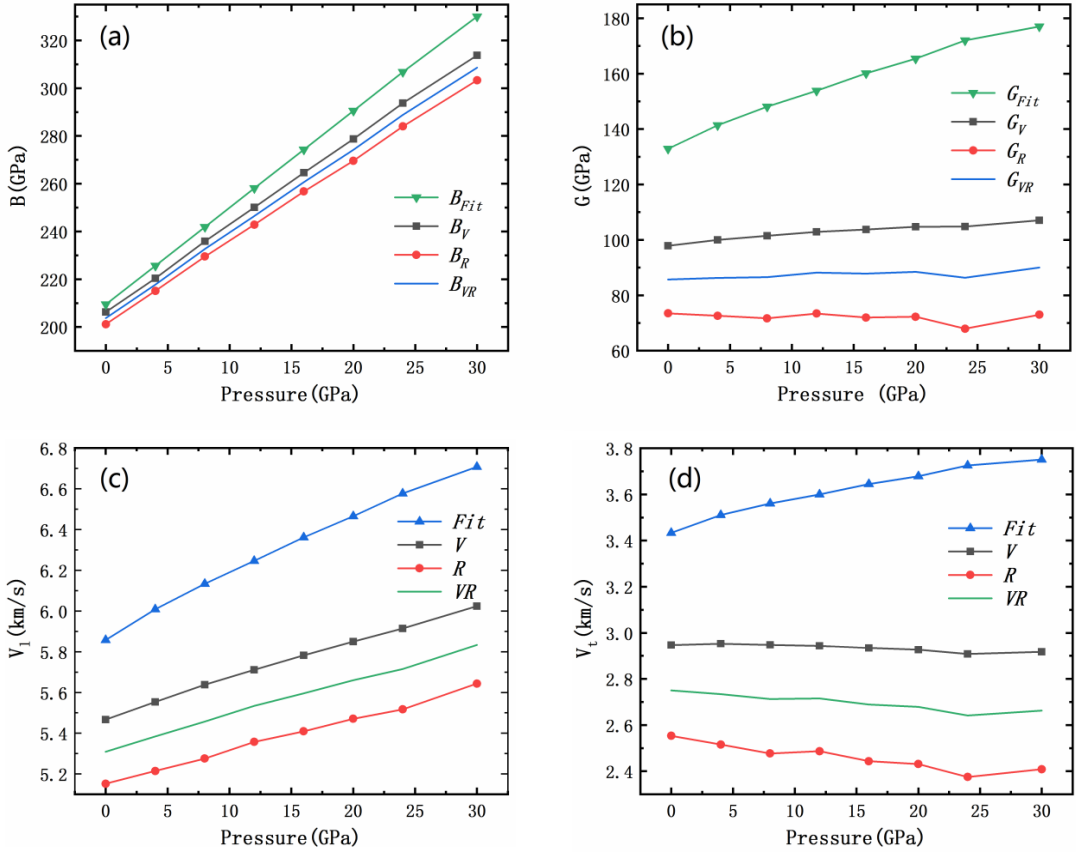

**Fig. 8** **(a)** Bulk modulus, **(b)** shear modulus, **(c)** longitudinal sound velocity and **(d)** transverse sound velocity at different pressures. The subscript *Fit* represents the moduli obtained by fitting Eq. (1) and Eqs. (7)-(9) using the data of specific heat, and the subscripts *V*, *R* and *VR* represent Voigt and Reuss approximation and their arithmetic average [Eq. (5)], respectively.

As shown in Fig. 8(b), the effective shear moduli ($G_V$ and $G_R$) vary much more slowly with pressure than the bulk moduli ($B_V$, $B_R$). The pressure derivatives of the average bulk and shear moduli were 3.54 and 0.07, respectively. Additionally, the pressure variations in the longitudinal and transverse sound velocities were ~ 17 m/s·GPa$^{-1}$ and −3.4 m/s·GPa$^{-1}$, respectively. The nearly constant shear moduli imply that the transverse sound velocity is nearly invariant within the pressure range 0–30 GPa [as shown in Fig. 8(d)]. Furthermore, we considered the effectiveness of the different pseudopotentials. The pressure variation of the elastic modulus obtained using the hard pseudopotential was consistent with that of the standard DFT–GGA calculations (the two results deviated within 5%). Therefore, standard GGA potentials are still valid for describing the $β$-PtO$_2$ system under high pressure (~30 GPa).

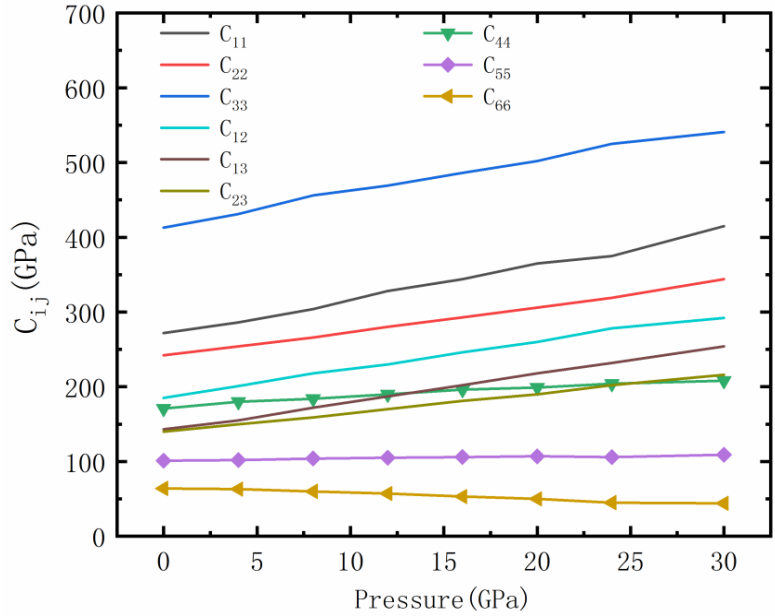

**Fig. 9** Pressure dependence of the elastic stiffness coefficient $C_{ij}$.

To explore the fundamental cause of the slowly changing shear modulus, we plotted the elastic stiffness coefficients as functions of pressure (Fig. 9). We observe that $C_{66}$ decreases with increasing pressure and that $C_{44}$, $C_{55}$, and $C_{66}$ change much more slowly than the other stiffness coefficients. Combined with the expression of the shear modulus [Eq. (5)], we infer that the slow variation of the shear modulus is mainly owing to the three stiffness coefficients, $C_{44}$, $C_{55}$, and $C_{66}$, which describe the weak shear motions of the *xy*, *xz*, and *yz* planes of the orthorhombic $β$-PtO$_2$. To our knowledge, the anomalies of the elastic coefficients $C_{44}$, $C_{55}$, and $C_{66}$ in transition metal dioxide (TMO$_2$) systems have not been previously reported; however, such anomalies in cubic lattice materials (e.g., ZnAl$_2$O$_4$ and Fe$_3$O$_4$) have been reported. In these systems, the elastic coefficient $C_{44}$ changes slowly or decreases with pressure [38, 39]. Such anomalous behavior occurs because of the relationship between the shear modulus and sliding moments at the grain boundaries of polycrystalline materials, which are less sensitive to volume compression of the composite crystallites. Therefore, we expect that the anomaly of shear modulus would also exist in the other systems such as tetragonal and hexagonal polycrystalline materials.

## 4 Conclusions

Based on the DFT calculations, we have systematically studied the structure and phononic, thermodynamic, and elastic properties of $\beta$-PtO$_2$ under different pressures. The calculated lattice parameters are in good agreement with those reported in previous works. An extensive investigation is presented on the characteristics of vibrational modes of $\beta$-PtO$_2$, which is found to be stable over a wide range of pressures. The phonon modes of $\beta$-PtO$_2$ in the high-frequency region were discretized and enhanced under pressure. The specific heat was confirmed to follow the Debye $T^3$ law. The Debye temperatures were deduced from the specific-heat data (single-crystal) and macroscopic effective elastic constants. The two sets of results were compared in the pressure range 0–30 GPa. As evidenced by the notable difference between the calculated Debye temperatures, the results of anisotropic single-crystal systems are not applicable in isotropic polycrystalline systems. The pressure dependences of the effective elastic moduli were calculated and compared through the Voigt–Reuss–Hill approximation. The bulk and shear moduli responded differently to increasing pressure: the bulk modulus increased uniformly whereas the shear modulus varied slowly. The slow variation of the shear modulus causes the transverse sound velocity to be nearly independent of pressure variations. We expect that these results can be further tested in future experimental studies.

**Acknowledgements** This work was supported by the National Natural Science Foundation of China (Grant No. 11474285). We gratefully acknowledge the high-performance supercomputing service from AM-HPC and the staff of the Hefei Branch of Supercomputing Center of Chinese Academy of Sciences for supporting the supercomputing facilities.